\documentclass[aip,rsi,reprint, amsmath,amssymb,]{revtex4-1}

\usepackage{graphicx}
\usepackage{bm}
\usepackage{dcolumn}

\begin{document}

\title{An in-vacuum diffractometer for resonant elastic soft x-ray scattering}

\author{D. G. Hawthorn}
\affiliation{Department of Physics and Astronomy, University of Waterloo, Waterloo, N2L 3G1, Canada}

\author{F. He}
\affiliation{Canadian Light Source, University of Saskatchewan, Saskatoon, Saskatchewan, S7N 0X4, Canada}

\author{L. Venema}
\affiliation{University of Groningen, 9747 AG Groningen, The Netherlands}

\author{H. Davis}
\affiliation{Department of Physics and Astronomy, University of British Columbia, Vancouver, V6T 1Z4, Canada}

\author{A. J. Achkar}
\affiliation{Department of Physics and Astronomy, University of Waterloo, Waterloo, N2L 3G1, Canada}

\author{J. Zhang}
\affiliation{Department of Physics and Astronomy, University of Waterloo, Waterloo, N2L 3G1, Canada}

\author{R. Sutarto}
\affiliation{Department of Physics and Astronomy, University of British Columbia, Vancouver, V6T 1Z4, Canada}

\author{H. Wadati}
\affiliation{Department of Applied Physics and Quantum-Phase Electronics Center (QPEC), University of Tokyo, Hongo, Tokyo 113-8656, Japan}
\affiliation{Department of Physics and Astronomy, University of British Columbia, Vancouver, V6T 1Z4, Canada}

\author{A. Radi}
\affiliation{Department of Chemistry, University of British Columbia, Vancouver, V6T 1Z4, Canada}

\author{T. Wilson}
\affiliation{Canadian Light Source, University of Saskatchewan, Saskatoon, Saskatchewan, S7N 0X4, Canada}

\author{G. Wright}
\affiliation{Canadian Light Source, University of Saskatchewan, Saskatoon, Saskatchewan, S7N 0X4, Canada}

\author{K. M. Shen}
\affiliation{Laboratory of Atomic and Solid State Physics, Department of Physics, Cornell University, Ithaca NY 14853}

\author{J. Geck}
\affiliation{Leibniz Institute for Solid State and Materials Research IFW Dresden, Helmholtzstrasse 20, 01069 Dresden, Germany}


\author{H. Zhang}
\affiliation{Department of Physics, University of Toronto, 60 St. George Street, Toronto, Ontario M5S 1A7, Canada}

\author{V. Nov\'ak}
\affiliation{Institute of Physics ASCR v.v.i., Cukrovarnick\'a 10, 162 53 Praha 6, Czech Republic}

\author{G. A. Sawatzky}
\affiliation{Department of Physics and Astronomy, University of British Columbia, Vancouver, V6T 1Z4, Canada}
\affiliation{Department of Chemistry, University of British Columbia, Vancouver, V6T 1Z4, Canada}

\date{\today}

\begin{abstract}
We describe the design, construction and performance of a 4-circle in-vacuum diffractometer for resonant elastic soft x-ray scattering and reflectivity.   The diffractometer, installed on the REIXS beamline at the Canadian Light Source, includes 9 in-vacuum motions driven by in-vacuum stepper motors and operates in ultra-high vacuum at base pressure of 2$\times$10$^{-10}$ Torr.   Cooling to a base temperature of 18 K is provided with a closed-cycle cryostat.  The diffractometer includes a choice of 3 photon detectors: a photodiode, a channeltron and a 2D sensitive channelplate detector.  Along with variable slit and filter options, these detectors are suitable for studying a wide range of phenomena having both weak and strong diffraction signals.  Example measurements of diffraction and reflectivity in Nd-doped (La,Sr)$_2$CuO$_4$ and thin film (Ga,Mn)As are shown. \end{abstract}

\maketitle

\section{Introduction}
In the past several years, resonant elastic soft x-ray scattering (RSXS) has emerged as a probe with unique sensitivities to spin, charge and orbital order.\cite{Abbamonte04,Abbamonte05,Wilkins03,Thomas04,Schusler05,Huang06}  The key of the technique is to combine x-ray spectroscopy, a probe of electronic structure, with x-ray diffraction, a probe of spatial order.  By tuning the energy of an x-ray to specific x-ray absorption edges, x-ray scattering can be made very sensitive to particular atomic orbitals of an element.  This gives RSXS not only valuable sensitivity to specific elements but also sensitivity to specific atomic orbitals (e.g. O $2p$ or Cu $3d$ states) of a given element as well as the spin and symmetry of those orbitals (e.g. the projected O $2p_x$ states).    By working in the soft x-ray region one can tune the photon energy to the x-ray absorption edges that correspond to the electronic states that are most important for the low energy physics of the materials, such as transition metal oxides, we wish to study (the O, N or C, $2p$ states, the transition metal $3d$ states and the rare-earth $4f$ states).  This allows RSXS to be used to study and distinguish between a wide variety of spin, charge, orbital and lattice order very directly and in considerable detail.

While powerful, technically, it is a challenging task to construct a reliable RSXS setup.  The attenuation length of soft x-rays in air and in the target material is much shorter than hard x-rays.  This necessitates high vacuum or ideally ultra high vacuum (UHV) conditions, ruling out much of the technology used in hard x-ray diffractometers.  This can be accomplished by coupling motions external to the vacuum into a vacuum chamber via differentially pumped rotary feedthroughs.  Several soft x-ray diffractometers have been designed along this line.\cite{Beale10, Bruck08, Grabis03, Roper01,Yu05, Takeuchi09}  In this manuscript we describe a soft x-ray diffractometer using in-vacuum stepper motors, with the majority of the moving parts in UHV.

This endstation is permanently installed at the Resonant Elastic and Inelastic X-ray Scattering (REIXS) beamline of the Canadian Light Source (CLS).   This beamline is an EPU beamline providing full control over the incident polarization.  The beamline covers an energy range 80 - 2500 eV, has an energy resolution $\Delta E/E$ = 10$^{-4}$ at 1000 eV, a flux of 5 $\times 10^{12}$ photons/s/0.1\% BW (at 1000 eV and 100 mA ring current) and a spot size of $\sim$250 $\mu$m (horizontal) $\times \sim$150 $\mu$m (vertical) at the sample position.  

\section{Diffractometer Design}

\begin{figure}[tbp]
\begin{center}
\resizebox{\columnwidth}{!}{\includegraphics{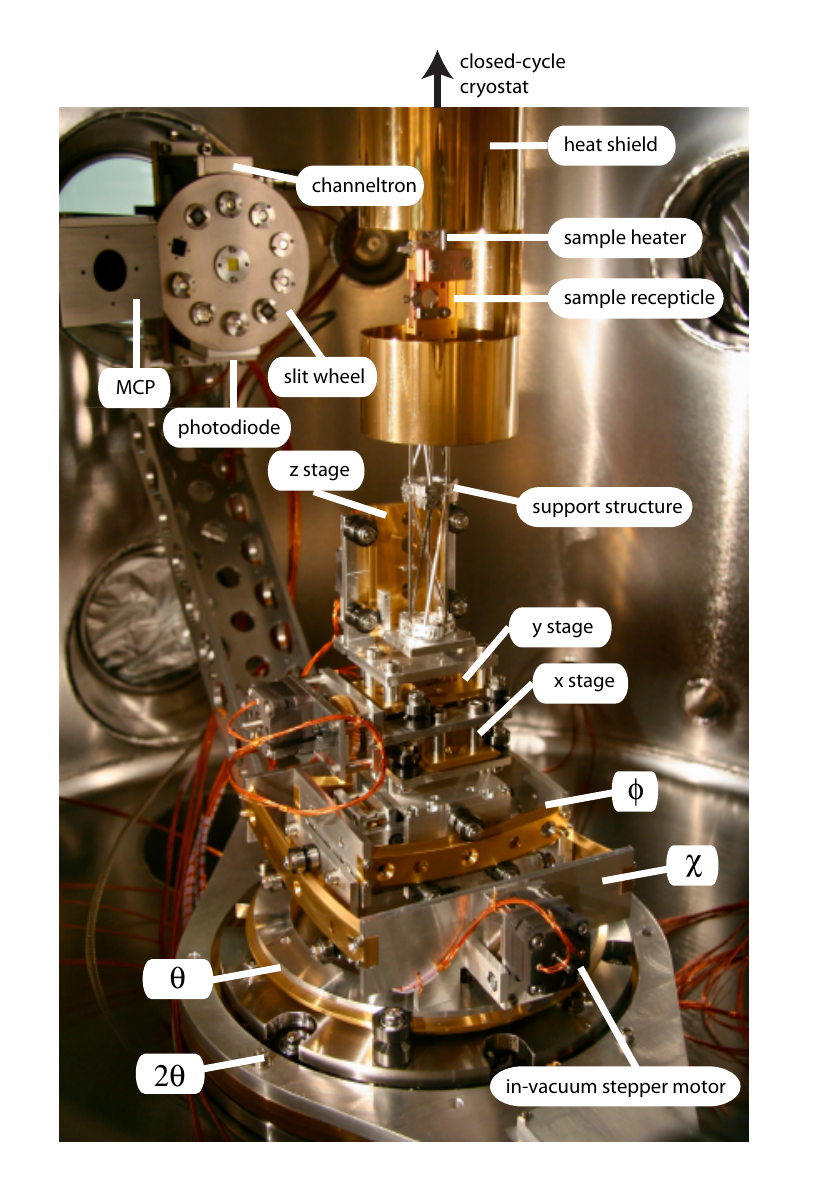}}
\caption{A photo of the installed RSXS diffractometer.}
\label{fig:photo}
\end{center}
\end{figure}

The diffractometer, shown in figure~\ref{fig:photo}, is a 4-circle UHV diffractometer with a total of 10 degrees of motion based in part on the ``Spinoza diffractometer'' at the National Synchrotron Light Source (NSLS), beamline X1B.  The diffractometer has a full range of $\theta$ and 2$\theta$ motions, spanning -25 to +265 degrees, and limited $\chi$ and $\phi$ motions, each spanning $\pm$ 4.5 degrees.  In addition, the sample position can be moved $\pm$ 7.5 mm in $x$, $y$ and $z$ to position a sample into the center of rotation of the diffractometer.  

In addition to these 7 motions, the detector arm allows for a height adjustment of 90 mm to position any of the three detectors (photodiode, channeltron or MCP) into the horizontal scattering plane.  A slit wheel with unlimited 360 degree rotation allows for one of ten filters/slits to be positioned in front of either the channeltron or photodiode detector.  All of these motions are achieved using in-vacuum stepper motors (Arun Microelectronics Ltd. C14.1 motors) and suitable gear reduction.  In addition to the in-vacuum motions, the cryostat mounted on the vacuum chamber can be rotated via a differentially pumped rotary feedthrough. This is motorized externally and allows for the cryostat to be rotated in concert with or separately from the $\theta$ rotation.  The motors are driven by Parker Hannifin Corp. E-DC and OrientalMotor RBD245A-V stepper motor drivers and Oregon Micro Systems MaxV motor controllers using both full step and micro-stepping modes depending on the motion.

The in-vacuum motions are provided by custom gearing, bearings and v-groove rollers with gear reduction ranging from 1:2 for the linear motions to 1:1800 for the $\theta$ and 2$\theta$ motions.  Worm gears are used for the $\theta$ and 2$\theta$ motions in order to reduce backlash and achieve the high gearing ratio (high torque).  This gearing provides precision of less than $0.001^\circ$ for $\theta$ and 2$\theta$, 2.5 $\mu$m for the linear motions ($x$, $y$, $z$, detector height) and 0.0005$^\circ$ for the $\phi$ and $\chi$ motions.  To reduce wear and seizing, the gears and v-groove rollers are coated with either diamond-like carbon (DLC), TiN or MoS$_2$ coatings.  In addition, WS$_2$ powder was applied as a dry lubricant on all gears.

The diffractometer is mounted on a subframe to provide manual, out-of-vacuum adjustments for the pitch, yaw, vertical height and horizontal position of the diffractometer relative to the beam direction.   This allows for orientation of the diffractometer to provide a horizontal scattering plane with the beam intersecting the axis of rotation of the goniometer.  The subframe is connected to the vacuum chamber via edge-welded bellows and can be adjusted independently of the scattering chamber, which is mounted to the floor by a separate frame.    The flexible bellows also provides vibration isolation between the diffractometer and chamber, which is attached directly to sources of vibration such as the cryostat and pumps.

The experiment is controlled by spec x-ray diffraction and data acquisition software (Certified Scientific Software), which interfaces to EPICS based device drivers that control the detection electronics, beamline and motion control.

\subsection{Sample holder and cryostat}

\begin{figure}[tbp]
\begin{center}
\resizebox{\columnwidth}{!}{\includegraphics{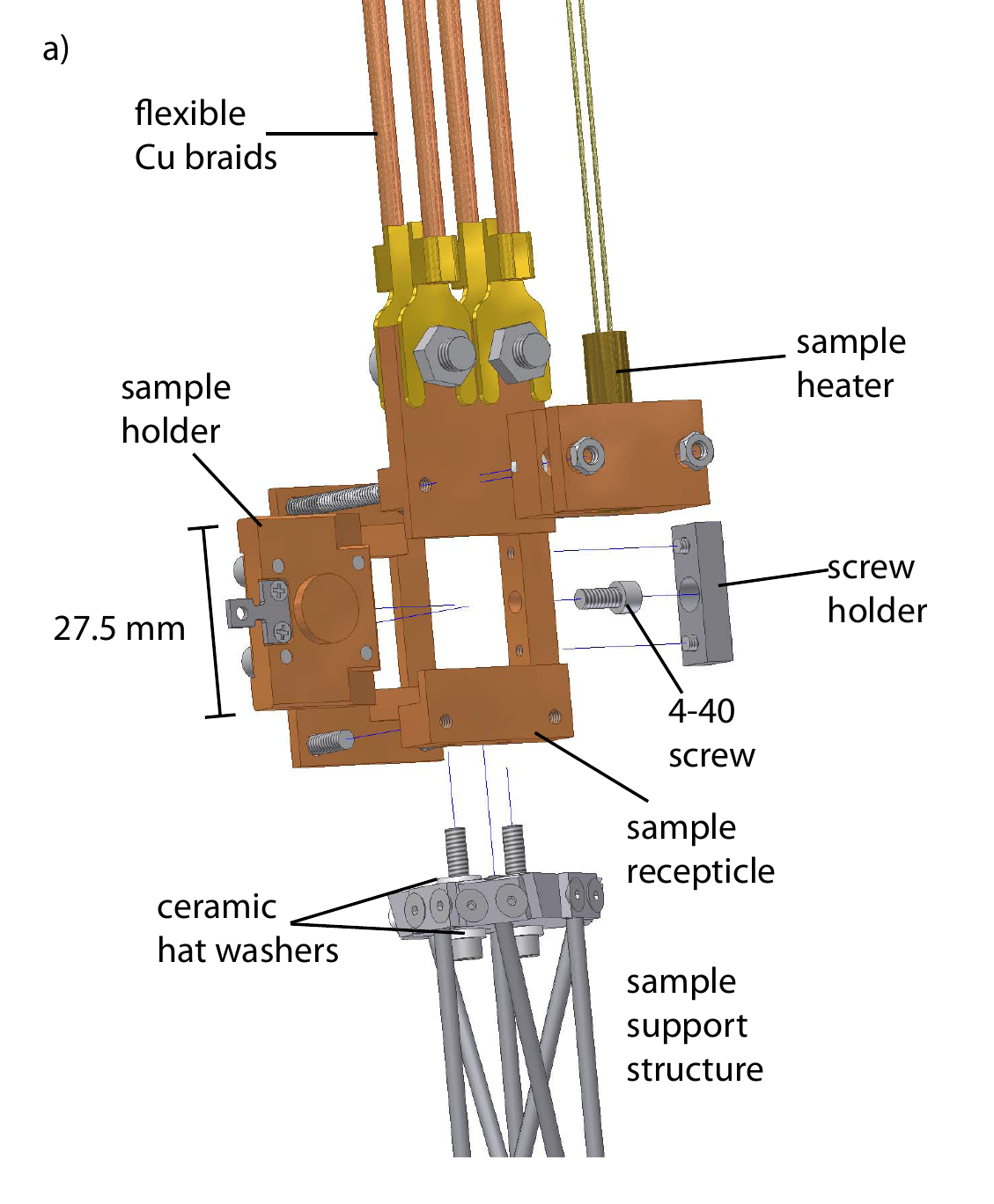}}
\caption{An expanded drawing of the sample holder and sample receptacle assembly.}
\label{sampleholder1}
\end{center}
\end{figure}

The sample holder and sample receptacle are shown in Fig.~\ref{sampleholder1}.  The sample holders have a wedge-shape that mates with a matching wedge-shaped receptacle.  With this design the sample holder fits loosely into the sample receptacle and the wedge-shaped sample holder provides a self-guided alignment into the slot of the sample holder during sample transfer.  In order to fix the sample holder into the sample receptacle, a \#4-40 socket head cap screw that is fixed to the sample receptacle (but free to rotate) is used to lock into a  \#4-40 thread on the sample holder and draw the sample holder towards the receptacle.   Contact is achieved at the wedges of the sample holder, providing good thermal and electrical contact.  The screw also provides active release of the sample holder, such that little force is required to remove the sample holder from the sample receptacle.  The screw action is provided by a Ferrovac WM40 magnetically coupled wobblestick manipulator with 3/32 inch ball-head hex tool attachment.  

Samples can be transferred in and out the the scattering chamber via a load-lock and magnetically coupled sample transporter (Ferrovac RMDG).  The load lock has a garage that can accommodate 3 sample positions that can be moved out of the path of the sample transfer arm using a rack-and-pinion linear translator.  The sample transporter has a Ferrovac pincer grip sample grabber that is used to clamp around the ``Omicron-style'' stainless steel eyehole protrusion on the sample holder.  When the pincer grip is closed around the eyehole protrusion, it fits safely in place with little chance of inadvertently dropping a sample in the chamber.  The load-lock is also equipped with a Kratos blade-and-anvil style sample cleaver, which can be used to cleave or fracture samples in the load lock at a typical pressure of 2 $\times$ 10$^{-8}$ Torr prior to transfer.

Two types of sample holder have been developed for the RSXS endstation; a plain flat sample holder and a sample holder that accommodates a cylindrical plug.  This latter sample holder allows for continuous in-vacuum azimuthal rotation of the sample by first loosing 4 clamping screws using the in-vacuum screwdriver, rotating the cylindrical sample plug (which has a \#4-40 screw in the back) using the screwdriver, and then retightening the clamping screws.  This feature expands the $\phi$ rotation range from the motorized $\pm$4.5$^\circ$ to a full 360$^\circ$ rotation.  Alternately, these latter sample holders can accommodate disk-shaped, 1/2 inch diameter, NdFeB permanent magnets, in order to perform measurements of samples in a fixed magnetic field.  For these in-field measurements, the sample is typically mounted directly on the permanent magnet and the additional in-vacuum azimuthal rotation of the sample is not provided.

From above, the sample receptacle is connected to the cryostat using 4 lengths of $1/8$'' OD OFHC copper rope with gold-plated copper lugs clamped to each end of the rope.  On the cryostat end, the copper rope is screwed down to an adapter that is separated from the cryostat by a thin sapphire plate to provide electrical isolation but maintain good thermal contact.

The sample receptacle is attached to the diffractometer via a thermally isolating support structure constructed out of thin wall stainless steel hypodermic tubing (0.095'' diameter, 0.005'' thick).  This structure has two levels with 6 tubes per level, held together by aluminum brackets.  The sample receptacle is electrically isolated from the support structure by ceramic hat washers.  This design provides high strength and stability with minimal conductive heat loss through the support structure. 

The sample receptacle, copper braid clamps and some of the sample holders were electropolished and electroplated with 2 $\mu$m Ni followed by 5 $\mu$m Au in order to reduce their thermal emissivity.

The sample is cooled with an Advanced Research Systems (ARS) DE-210SB closed cycle cryostat.  The cryostat is a UHV compatible, 2-stage cryostat.  The 2nd stage of the cryostat is designed to reach a base temperature $<3$ K with a full heat shield and have 0.8 W of cooling power at 4.2 K.  A gold-plated, OFHC copper heat shield is attached to the first stage of the cryostat and extends to partially cover the sample holder in an effort to reduce radiative heat losses.  At base temperature, the first stage of the cryostat is cooled to 9K and the sample is at 18 K.  These temperatures are limited by the radiative heat losses through the openings in the heat-shield (The heat load on the sample supplied by the support structure is estimated to be 0.1W).  A planned modification of the heat shield to extend the coverage are expected to further reduce the base temperature at the sample position.  

The temperature of the cryostat is monitored and controlled using three silicon diode temperature sensors (Lakeshore DT-670B-SD) with one sensor close to the sample position, a second sensor on the second stage of the cryostat and a third sensor intermediate between the two.  The sample is heated using a 25 $\Omega$, 100 W cartridge heater (Lakeshore HTR-25-100) located just above the sample position.  This heater is thermally connected but electrically isolated from the sample receptacle using a sapphire plate and ceramic hat washers.  This is done in order to eliminate leakage currents from the heater influencing total electron yield drain current measurements.  This configuration of the heater allows the sample temperature to be raised from a base temperature of 18 K up to 300 K while maintaining the 2nd stage of the cryostat below 30 K.

An appealing feature of this design is that thermal expansion and contraction is minimal.  Inspection of the sample position using a telescope while cooling between room temperature and base temperature indicate the change in the vertical height of the sample is less than 100 $\mu$m.  Large contractions can be avoided because the cryostat is mechanically isolated from the sample via the copper braids and only the shorter sample stage and support structure expand or contract upon heating or cooling.  This feature definitely reduces the need to reposition the sample in the beam for scans at different temperatures. 

\subsection{Detectors}

The diffractometer is equipped with 3 detectors: a photodiode, a channeltron and channelplate detector.  These detectors can be translated vertically by 9 cm to position any of the detectors into the horizontal scattering plane.  The photodiode detector has a large dynamic range and can be used for scattering, reflectivity and x-ray absorption measurements.  The channeltron detector is a single photon sensitive detector that can be used to measure weak superlattice peaks or x-ray fluorescence.  The MCP detector is also a single photon sensitive detector, like the channeltron, that has 2D spatial sensitivity and thus can be used for more rapid scanning of reciprocal space. 

The photodiode is an AXUV100EUT detector from International Radiation detectors Inc. and has a 10 mm $\times$ 10 mm  square detection area.  With the exception of a dark current of order 0.1 to 4 pA, the response of the photodiode is reported to be linear in photon intensity over a large dynamic range, extending from $\sim 10^{-5}$ A for the direct beam to currents in the Picoampere range for x-ray fluorescence.  This detector is enclosed in an aluminum enclosure and wired using triaxial cable and connectors to allow for a guarded and shielded measurement from the photodiode to the Keithley model 6514 electrometer, which measures the photocurrent.  For these measurements, the anode ($p$) of the photodiode is connected to the measurement HI and the cathode ($n$) is connected to the measurement LO of the electrometer.

The channeltron used is a Sjuts KBL1010.  When a photon or charged particle hits the channeltron, it produces secondary electrons that are accelerated by the high voltage applied to the channeltron.  These electrons will also strike the walls of the channeltron producing additional electrons that are further accelerated down the channeltron, creating an avalanche effect.  This avalanche process produces a short pulse ($\sim 8$ns) of up to $~10^8$ electrons at the anode per photon.  These pulses are counted to determine the number of incident particles hitting the channeltron.  The Sjuts KBL1010 channeltron has a maximum count rate of $\sim 10^6$ counts/s at which point the count rate will become non-linear in the incident photon intensity.  With no beam the dark counts on the detector are $<$ 0.1 counts/s.  The channeltron is used in a configuration where photons will be directly incident on the channeltron's surface.  In front of the channeltron is situated an electro-formed gold mesh with 85\% transmission for repulsion of electrons and ions.  This grid is electrically isolated from ground by ceramic standoffs.  

\begin{figure}[tbp]
\begin{center}
\resizebox{\columnwidth}{!}{\includegraphics{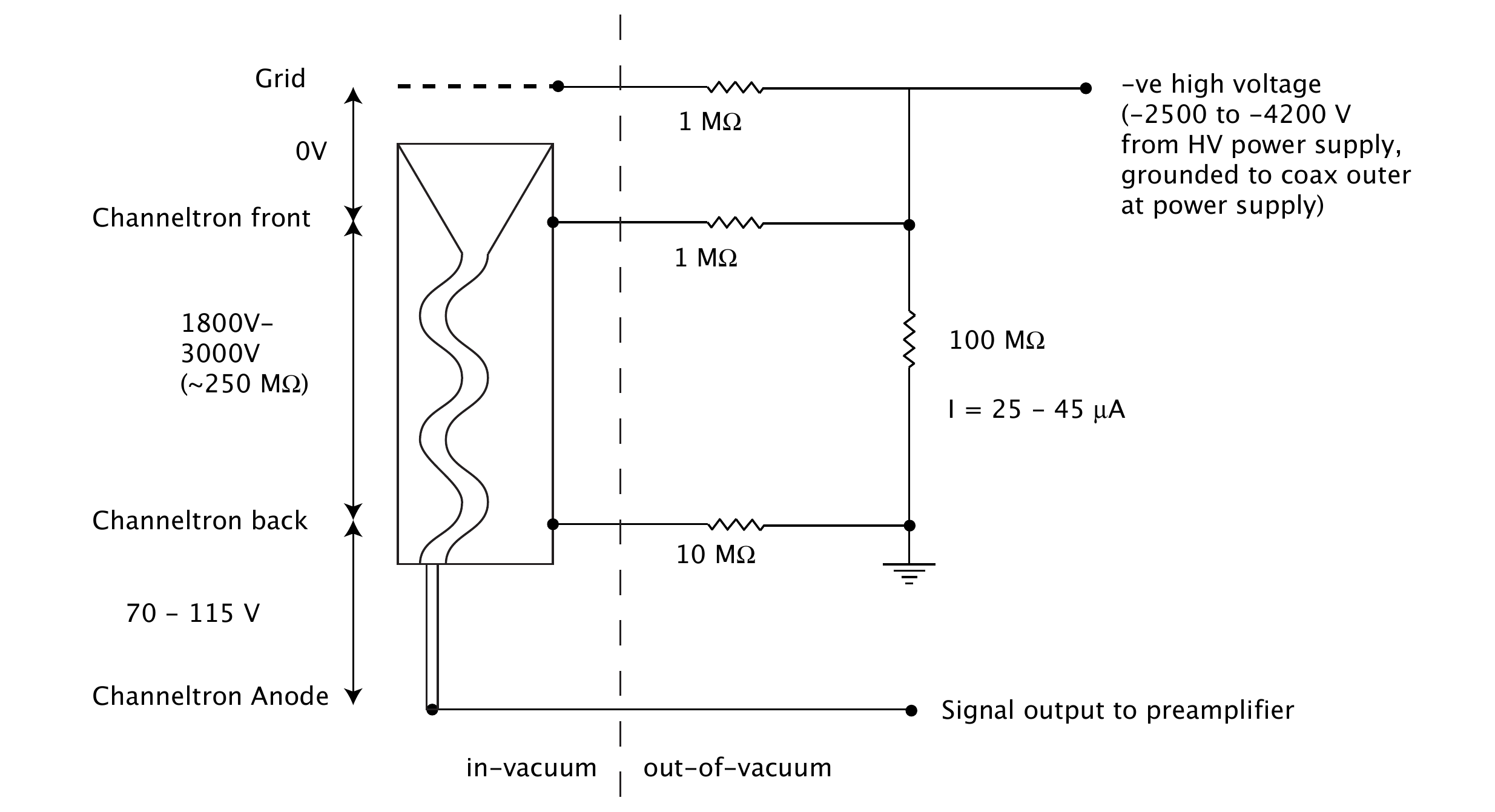}}
\caption{Voltage divider for -ve biasing of the channeltron.}
\label{chanHV-}
\end{center}
\end{figure}

The channeltron can be positively or negatively biased by applying appropriate biasing to the grid, the channeltron front and the channeltron back.  This is accomplished using a high voltage power supply and the appropriate voltage divider.  In most instances, a negative bias configuration is employed.  With a negative bias, the anode of the channeltron is at ground and the front end of the channeltron is held at a large negative voltage.  This biasing configuration serves to repel negatively charged ions or electrons from the front end of the channeltron.  Positive ions in the chamber, produced from the ion gauges or from the x-ray beam hitting the sample, will be attracted to the channeltron.  Some of these ions will be intercepted by the grid in front of the channeltron, which is also held at a large negative bias.  The back of the channeltron is held at $\sim100$ V above the channeltron anode to ensure that electrons in the channeltron are accelerated to the anode rather than the back of the channeltron.  The voltage divider shown in figure \ref{chanHV-} can be used for biasing with a single high-voltage power supply.   The channeltron is housed in the scattering chamber within an aluminum enclosure.  In-vacuum electrical connections to the channeltron are made using MHV panel jack receptacles.  These receptacles are connected to SHV vacuum feedthroughs using Accuglass kapton insulated co-axial cables that have MHV connectors on one end and Accufast 500s connectors on the other end.  To count pulses from the channeltron, the charge pulse is first amplified by an Ortec VT120C high speed voltage preamplifier that has a gain of 20 and a 50 $\Omega$ input impedance.  The preamplifier also provides appropriate impedance matching to the Ortec 584 constant fraction discriminator.  The output of the discriminator is counted by a Struck Innovation Systeme SIS3820-64M scaler card and simultaneously monitored with an Ortec 449-2 ratemeter.

In front of the channeltron and photodiode detectors is a slit wheel that can position one of ten slit/filter combinations in front of either the channeltron or photodiode detectors.  The slits and filters are mounted on modular stainless steel rings that can be screwed into the slit wheel.  This modular design allows for easy replacement of filters and for slits and filters to be installed in various combinations.  The present slit options are 10 mm $\times$ 10 mm, 1 mm $\times$ 2 mm, 0.5 mm $\times$ 3 mm, 0.1 mm $\times$ 3 mm and 0.1 mm $\times$ 1 mm and are oriented such that the long axis of the slit will be vertical (short axis in the scattering plane) when the slits are positioned in front of the channeltron or photodiode.  With a sample-to-detector distance of 290 mm,  the 0.5 mm slit corresponds to 2$\theta$ angular width of 0.1$^\circ$.

\begin{figure}[tbp]
\begin{center}
\resizebox{\columnwidth}{!}{\includegraphics{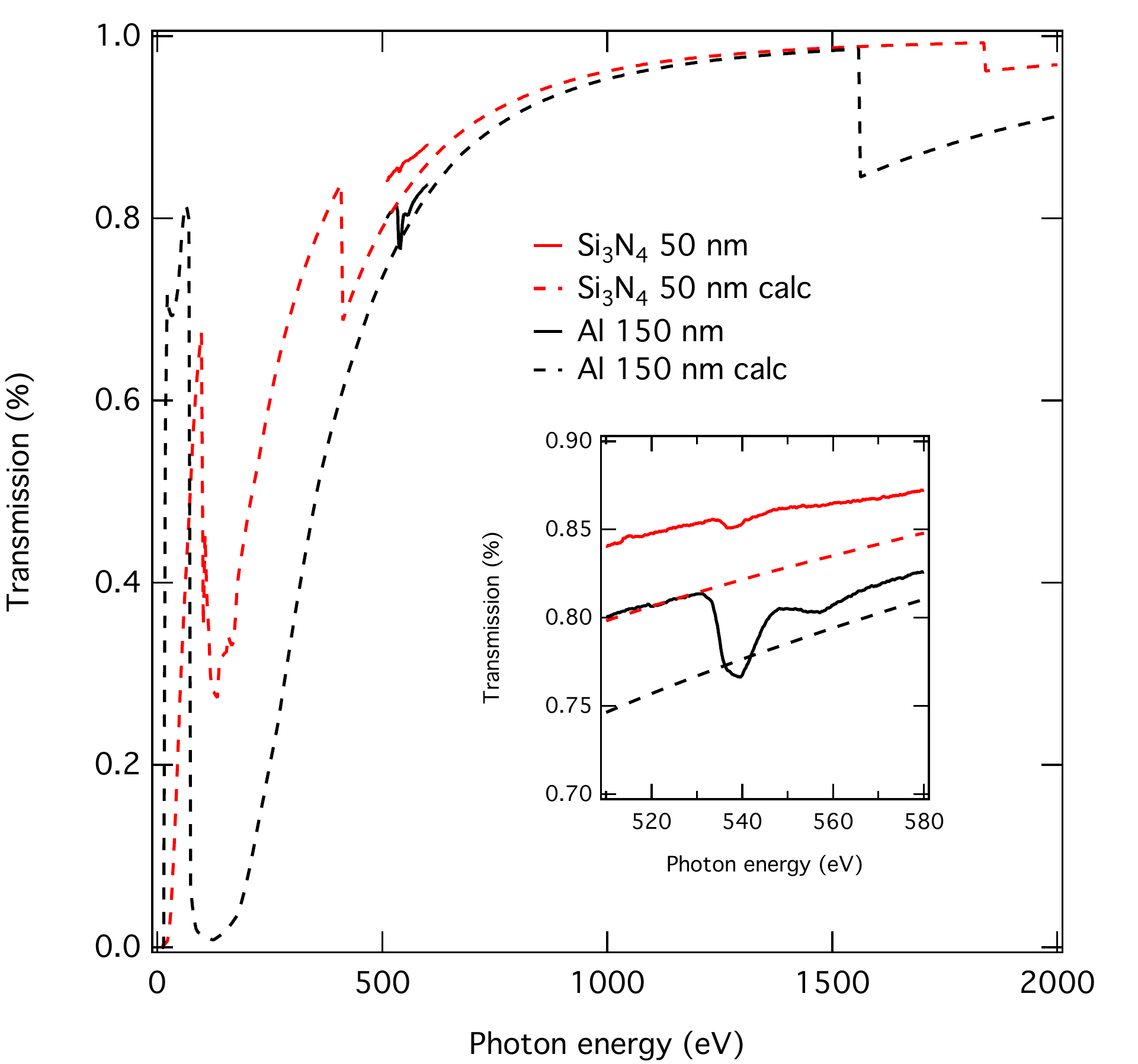}}
\caption{Measured (solid) and calculated (hashed) transmission vs. photon energy for the 150 nm thick aluminum (black) and 50 nm thick silicon nitride (red)  filters.   The calculated transmission is done using tabulated values from Henke {\it et al.}\cite{Henke93}.  The inset shows a blowup of transmission through the O $K$ edge.  The Al filter shows some signs of oxidation, evidenced by an additional 7\% dip in transmission at the O $K$ edge.}
\label{Alfilter}
\end{center}
\end{figure}

In addition to the slits, the slit wheel can be used to position thin Al or SiN filters in front of either the channeltron or photodiode detectors.  The filters are used to block stray charged particles (electrons or positive ions) from the detectors, both of which are sensitive to charged particles as well as photons.  The Al filters are also particularly useful for blocking visible light from the photodiode detector, which may leak in through the beamline or from poorly covered windows and provide unwanted offsets in the x-ray induced photocurrent.  In addition, the filters will reduce contributions from low energy fluorescence to the background, which can enhance the elastically scattered signal relative to low energy fluorescent background in some instances.  The aluminum filters from Lebow company are 0.15 $\mu m$ thick foil and either 10 mm diameter or 5 mm diameter.  The silicon nitride (Si$_3$N$_4$) windows (Norcada, part \# NX10500A) have a 5 mm x 5 mm window that is 50 nm thick mounted on a 10 mm $\times$ 10 mm, 200 micron thick silicon frame.  This frame is then mounted on a Lebow LS0.5 or LS1.0 stainless steel ring.  The transmission vs. photon energy for the filters, determined from Ref.~\onlinecite{Henke93}, is shown in figure~\ref{Alfilter} along with measurements of the filter transmission around the O $K$ edge.  

The micro-channelplate (MCP) detector is a single photon sensitive, spatially resolved 2D detector from Quantar Technology Inc.  The channelplate detector head is a 25 mm diameter, double channelplate detector (part \# 3390-3090-SH-SM-R1).  This detector head has a lower resistance (faster time constant) resistive anode and is equipped with higher count-rate channelplates.  It has a spatial resolution of $\sim$0.25 mm and a maximum count rate of  $\sim10^6$ cts/s, at which point the count rate becomes non-linear with respect to photon intensity.  The detector head has a 85\% transmission electro-formed gold mesh attached to the front end.  The MCP is also coated in CsI in order to enhance its efficiency in the soft x-ray region.  It is mounted in an aluminum enclosure and attached to the enclosure by the grid ring (at the front of the detector).  This grounds the grid in front of the channelplate to the aluminum enclosure and subsequently the scattering chamber.  Similar to the channeltron, the channelplate requires high-voltage biasing in order to operate and is operated in a negative bias configuration so as to repel electrons.  This biasing is done using a single high voltage power supply and the voltage divider.  Like the channeltron, the in-vacuum high voltage electrical connections to the channelplate are made using MHV panel jack receptacles.  The signal leads use SMA panel jack receptacles.  These receptacles are connected to SHV or BNC vacuum feedthroughs using Accuglass kapton insulated co-axial cables.  The spatially resolved counts are read-out using a Quantar model 2502A position analyzer and model 2500 preamp.  

In addition to the photodiode, channeltron and channelplate detectors, the diffractometer is equipped to simultaneously measure the total electron yield (TEY) by measuring the drain current to the sample.  The drain current is measured using a Keithley picoammeter or a SRS570 current amplifier, the output of which can be converted to pulses for pulse counting using a Nova N101VTF voltage-to-frequency converter.  

For normalization of all measurements, the incident beam current, $I_0$, is measured in a similar fashion to the TEY, by measuring the drain current from a 85\% transmission gold mesh located between the sample and the last optical element in the beamline.

Finally, for a visible indication of the beam position, scintillating Yttrium aluminum garnet activated by cerium (YAG:Ce) crystals from Crytur Ltd. are installed on the detector arm (in the centre of the slit wheel) and as an optional sample when placed on the sample holder.  The YAG:Ce crystals are 10 mm $\times$ 10 mm $\times$ 1 mm with a square grid marked on the surface of the crystal.  The 8 mm $\times$ 8 mm grid has 1 mm $\times$ 1mm squares (30 $\mu m$ thick lines) and a cross-hair through the centre of the grid formed by thicker (60 $\mu m$) lines.  This YAG crystal is primarily used for diffractometer alignment, taking the role of burn paper in hard x-ray scattering measurements.  

\subsection{Detector Comparison}

Since the photodiode is an absolute detector with a known responsivity of 0.275 A/W at 640 eV (or 176 electrons/incident photon), the current of the photodiode can be used to calibrate the collection efficiency (number of x-rays detected/number of incident x-rays) of the other detectors by comparing fluorescence measurements with the detectors positioned at the same measurement position relative to the sample.  At 640 eV, this comparison (normalized to equal detector cross-sectional area) yields a collection efficiency of 27.5 \% for the CsI coated MCP and  $\sim 3 \%$ for the channeltron, compared with a 100\% collection efficiency for the photodiode detector.  However, given that the MCP detector has a cross-sectional area 4.9 times that of the photodiode, it provides the best signal-to-noise for fluorescence measurements.

For the channeltron and MCP detectors, the noise is dominated by $\sqrt{N}$ statistical noise.  However, for the photodiode detector, weak signals, such as the fluorescence measurement, are dominated by Johnson noise across the $\sim$ 100 M$\Omega$ shunt resistance of the photodiode.  At room temperature this gives a noise level of 13 fA/$\mathrm{\sqrt{Hz}}$.   This is the dominant source of noise for photocurrents less than 5 pA (a typical fluorescence signal is of order 1 pA).

In addition to the Johnson noise, the photodiode detector also has a dark current ranging from 0.1 to 4 pA.  The presence of the dark current in the absence of the beam on the sample may be related to the voltage burden of the electrometer across the 1M$\Omega$ shunt resistance of the photodiode.  However, it is likely that contributions to the dark noise also come from triboelectric or piezoelectric noise from the in-vacuum photodiode detector cable. 

\subsection{Vacuum system}

The scattering chamber is a $\sim1$m diameter  stainless steel vacuum chamber. The chamber is pumped by a Pfieffer 700 L/s turbopump and a CTI Cryotorr 8F cryopump.  The turbopump is backed by a triscroll pump that is used both to back the turbopump and to pump the chamber down from atmospheric pressure.  Both the turbopump and cryopump are separated from the scattering chamber by electro-pneumatic gate valves.  In addition, the scattering chamber has a differentially pumped rotary feedthrough, the first stage of which is pumped by the triscroll pump with the second stage pumped by a 2 L/s ion pump.  Pressure is measured by a combination of cold-cathode ion gauges, hot cathode ion gauges and thermocouple gauges.  During measurements, the hot cathode gauge, which is a source of both visible light and charged particles, is turned off.   A residual gas analyzer is also installed on the scattering chamber for vacuum characterization.  Measurements are performed at typical pressures of 2 - 6 $\times 10^{-10}$ Torr.

The scattering chamber is also designed to be used in conjunction with 3 other chambers: an oxide MBE chamber, a sample analysis chamber and a transfer chamber that serves as the hub of the three other chambers.  The combinations allows for the growth, characterization and measurement of samples without breaking vacuum.  The sample analysis chamber, designed by Omicron Inc., allows for UHV AFM/STM, EELS, UV photoemission, monochromatic x-ray photoemission, Auger spectroscopy, and SEM to be performed on samples prior to measurement in the scattering chamber.

When the scattering chamber is operated as a stand-alone system (without the sample prep and analysis chambers connected), the scattering chamber is connected directly to the load-lock chamber via a manual 4.5" CF gate valve.  The load-lock chamber is pumped by an 80 L/s turbopump that is backed by a single scroll pump.

\section{Results and discussion}  

\subsection{X-ray magnetic circular dichroism in Ni metal}
\begin{figure}[tbp]
\begin{center}
\resizebox{\columnwidth}{!}{\includegraphics{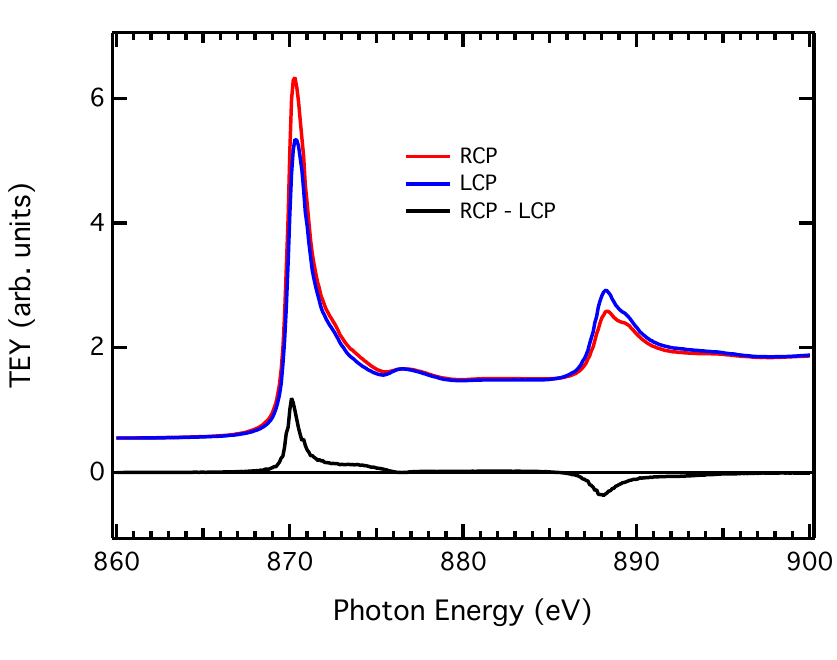}}
\caption{XAS and XMCD at the Ni $L$ edge in Ni metal measured using total electron yield.}
\label{NiXMCD}
\end{center}
\end{figure}

In figure~\ref{NiXMCD} we show x-ray magnetic circular dichroism measurements of Ni metal at the Ni $L$ edge.  The nickel sample was mounted directly on a NdFeB rare earth magnet with magnetic field perpendicular to the surface of the sample.  The surface of the Ni was scraped with a diamond file in vacuum prior to measurement.  The XMCD is measured using total electron yield with incident left circularly polarized (LCP) and right circularly polarized (RCP) light normal to the sample surface.  The TEY measurements shown in figure~\ref{NiXMCD} are in good agreement with results from literature,\cite{Chen90} demonstrating the control of polarization at the beamline.

\subsection{Charge stripe superlattice in La$_{1.475}$Nd$_{0.4}$Sr$_{0.125}$CuO$_4$}

As a test of the diffractometer for resonant soft x-ray diffraction, measurements were performed on the [0.25 0 $L$] superlattice peak in the stripe ordered cuprate  La$_{1.475}$Nd$_{0.4}$Sr$_{0.125}$CuO$_4$.\cite{Tranquada95}  Note we are referencing the Miller indices to the high-temperature tetragonal unit cell.  Similar to previous measurements of stripe ordering in La$_{1.875}$Ba$_{0.125}$CuO$_4$ and La$_{1.8-x}$Eu$_{0.2}$Sr$_{x}$CuO$_4$, \cite{Abbamonte05,Fink09} the stripe ordered superlattice peak in La$_{1.475}$Nd$_{0.4}$Sr$_{0.125}$CuO$_4$ is a weak peak that has a maximum intensity that is only a fraction of the fluorescent background, providing a good test of the ability to measure weak signals.

\begin{figure}[tbp]
\begin{center}
\resizebox{\columnwidth}{!}{\includegraphics{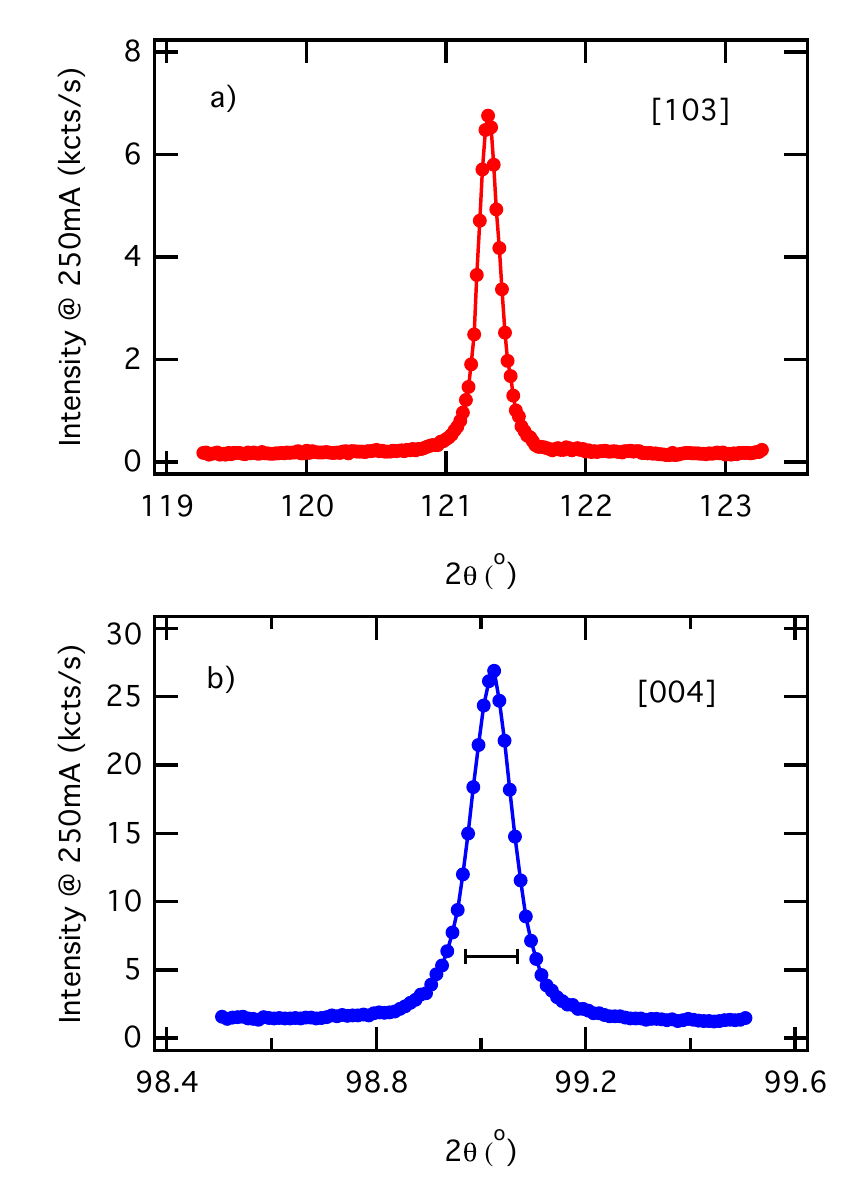}}
\caption{$\theta$-2$\theta$ scans through the [103] and [004] Bragg peaks of La$_{1.475}$Nd$_{0.4}$Sr$_{0.125}$CuO$_4$.  Both peaks were measured at 2491.8 eV using the channeltron detector.  The horizontal error bar indicates the width in angle of the 0.5 mm slit.  Intensity is plotted as kcts/s at 250 mA ring current.}
\label{figNdLSCOstruct-01}
\end{center}
\end{figure}

Prior to measuring the [0.25 0 $L$] superlattice peak, the sample is oriented using the [103] and [004] structural Bragg peaks measured at a photon energy of 2491.8 eV.  This photon energy is at the upper end of the capabilities of the REIXS beamline, leading to a photon intensity that is much less than the intensities achieved below 1000 eV.  Nonetheless, the photon intensity is sufficient to clearly measure structural Bragg peaks. This capability allows careful alignment of $\phi$ and $\chi$ prior to searching for weak superlattice peaks.  $\theta$ and $\theta$-2$\theta$ scans through these structural Bragg peaks are shown in figure~\ref{figNdLSCOstruct-01}.  These measurements were performed using the channeltron with a 0.5 mm $\times$ 3 mm (horizontal $\times$ vertical) slit and an aluminum filter.

\begin{figure}[tbp]
\begin{center}
\resizebox{\columnwidth}{!}{\includegraphics{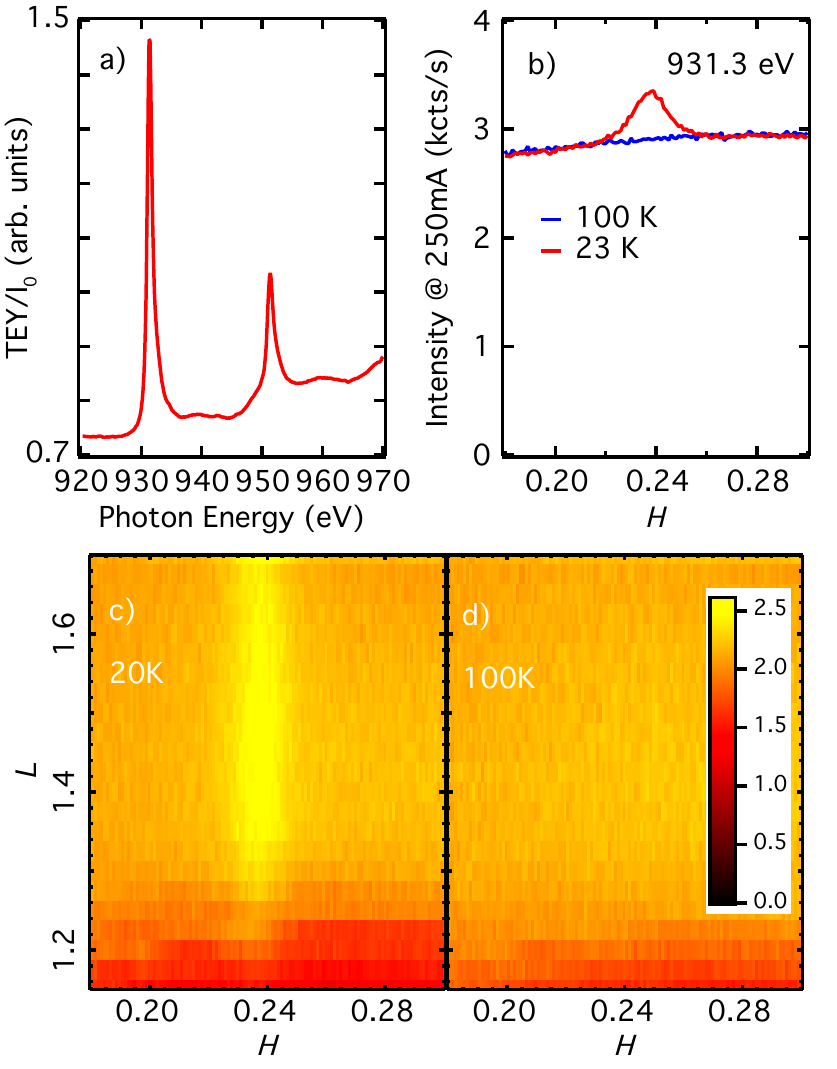}}
\caption{a)  XAS through the Cu $L$ edge in La$_{1.475}$Nd$_{0.4}$Sr$_{0.125}$CuO$_4$ measured using total electron yield.  b) Resonant scattering of the [$H$ 0 1.5] CDW superlattice peak measured at the peak of the Cu $L_3$ edge (931.3 eV).  Above the stripe ordering transition ($\simeq $70 K), a broad fluorescent background is observed.  Below the stripe ordering temperature, the CDW superlattice peak is observed above fluorescent background.  c) and d) Color maps of the photon intensity vs. $H$ and $L$ with $K$ = 0.  The CDW order forms a rod of scattering in $L$ centered on $H$ = 0.236.}
\label{figNdLSCOTdep-01}
\end{center}
\end{figure}

In figure~\ref{figNdLSCOTdep-01} a), the Cu $L$ edge XAS measured using TEY with the polarization along the $a$ axis of the sample is shown.  In order to expose a clean surface for the TEY measurement, the samples were cleaved perpendicular to the $c$-axis at a pressure of 2 $\times$ 10$^{-8}$ Torr using the Kratos crystal cleaver.  The lineshape of the XAS is consistent with previous measurements on La$_{2- x}$Sr$_{x}$CuO$_4$, \cite{Chen92} although the TEY signal is notably sharper due to the absence of self-absorption effects which are present in the fluorescence yield measurements of Chen {\it et al.}   In figure~\ref{figNdLSCOTdep-01} b), the scattering intensity through the [0.25 0 1.5] superlattice peak is shown for photon energy at the peak of the Cu $L_3$ absorption edge (931.3 eV).  At high temperatures, above the stripe ordering transition temperature ($T_s \simeq$75 K), a smoothly varying background is observed that is attributed to x-ray fluorescence.  At low temperature (23 K), the superlattice peak, centered at $H$ = 0.237, is clearly observed above the fluorescent background.  This peak has a peak position $H = 0.237$, consistent with neutron scattering measurements, \cite{Tranquada96} and peak width $\Delta H_{HWHM} = 0.008$ (correlation length = 73 \AA).  Notably, despite the weaker and broader superlattice peak in La$_{1.475}$Nd$_{0.4}$Sr$_{0.125}$CuO$_4$ relative to La$_{1.875}$Ba$_{0.125}$CuO$_4$ and La$_{1.8-x}$Eu$_{0.2}$Sr$_{x}$CuO$_4$, the signal-to-noise of our measurements compares favourably.

In figure~\ref{figNdLSCOTdep-01} c) and d), 2D maps of the measured intensity in $H$ and $L$ with an incident photon energy of 931.3 eV are shown.  Above the stripe ordering transition, a broad fluorescent background is observed that has an $H$ and $L$ dependence that is determined by the angle dependence of the fluorescence (the total or partial fluorescence yield depend on the angles of incidence and detection),\cite{Eisebitt93} which drops off at low values of $L$.  Below the stripe ordering transition, a rod of scattering is observed that is broad in $L$ and centred on $H = 0.236$, consistent with previous work on related materials.\cite{Abbamonte05}   

\subsection{Magnetic reflectivity in Ga$_{0.93}$Mn$_{0.07}$As}

The diffractometer can also be used for resonant reflectivity measurements of thin films and multilayers in addition to diffraction measurements.  Here we present test measurements of resonant magnetic reflectivity at the Mn $L$ edge of the dilute magnetic semiconductor Ga$_{0.93}$Mn$_{0.07}$As. This material is a 45 nm thick Ga$_{0.93}$Mn$_{0.07}$As film grown on the (001) surface of a GaAs substrate.  The sample measured was annealed at 200 $^{\circ}$C in air for 16 hrs.  Prior to measurement, the sample was etched in concentrated HCl in order to remove the surface oxide.\cite{Edmonds04}  The sample was mounted on a rare-earth permanent magnet with the magnetic field parallel to the [110] direction of the sample (parallel to the surface) and oriented in the horizontal scattering plane.  In this geometry the orientation of the magnetic field relative to the incident beam varies as a function of $\theta$.  Below the Curie temperature ($T_C$ = 155 K), Ga$_{0.93}$Mn$_{0.07}$As becomes ferromagnetic.  

\begin{figure}[tbp]
\begin{center}
\resizebox{\columnwidth}{!}{\includegraphics{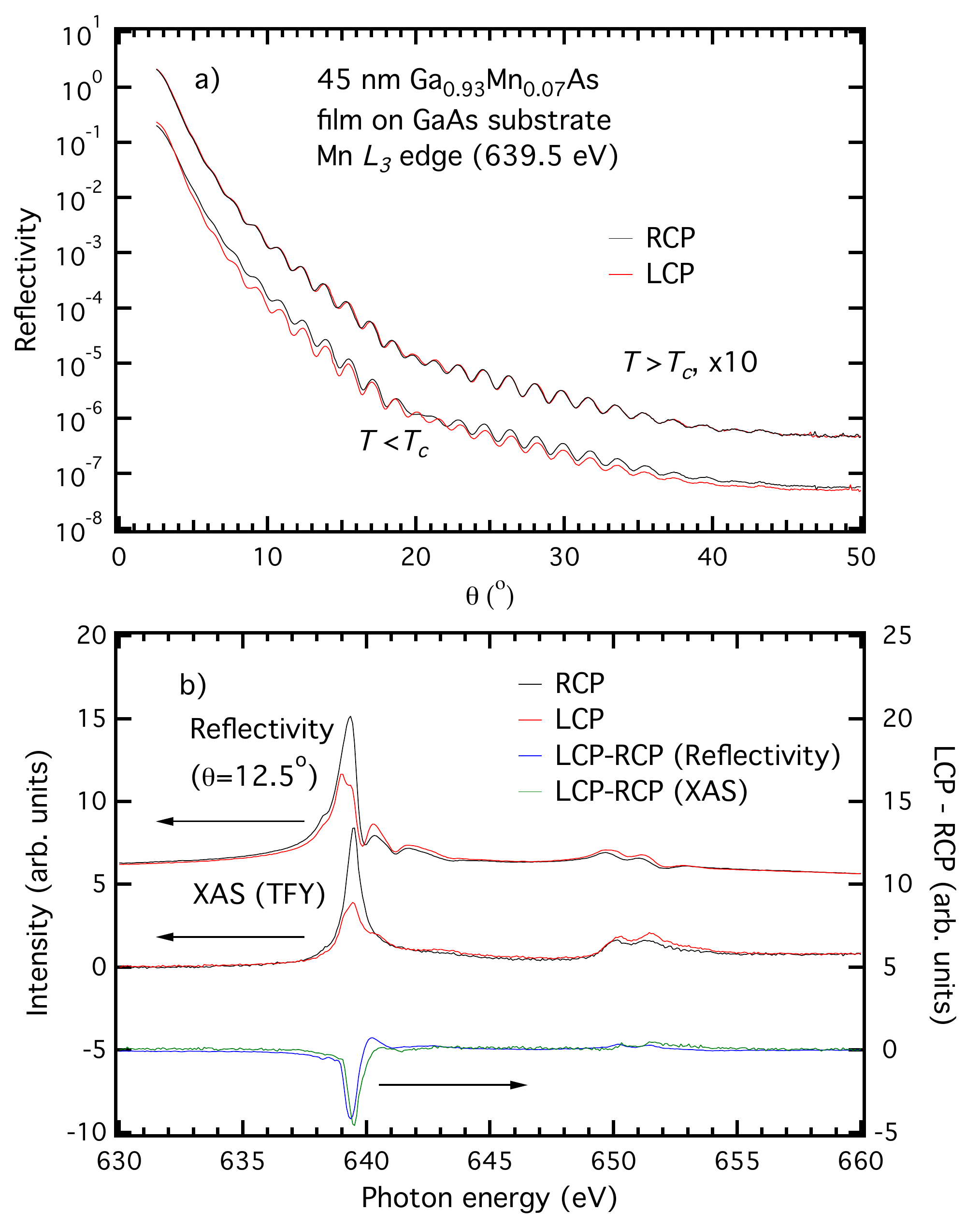}}
\caption{Magnetic resonant reflectivity measurements at the Mn $L$ edge of a thin film of Ga$_{0.93}$Mn$_{0.07}$As on a GaAs substrate.  a) Reflectivity vs. angle of incidence measured at the peak of the $L_3$ edge with left and right circularly polarized light.  The sample is measured both above and below the Curie temperature, with a dichroism observed only in the ferromagnetic state. b) Reflectivity with $\theta$ fixed at 12.5$^\circ$ and x-ray absorption as a function of photon energy. The x-ray absorption is measured using total fluorescence yield using the MCP detector.  The dichroism (LCP-RCP) of both the reflectivity and XAS have similar lineshapes and relative intensities.}
\label{fig:GMA}
\end{center}
\end{figure}

The reflectivity was measured with the photodiode detector using a 10 mm diameter slit and an aluminum filter window.  The dark current in these measurements was subtracted from the data prior to normalizing to $I_0$.  In figure~\ref{fig:GMA} a) we present magnetic resonant scattering of the (Ga,Mn)As film as a function of angle with the photon energy at the peak of the $L_3$ absorption edge (639.5 eV).  Here, the data is scaled to the direct beam current at 639.5 eV in order to provide reflectivity in the correct units ($R$ = 1 is perfect reflection).  Above $T_C$, the sample is paramagnetic and the measurements with left and right circularly polarized light overlap well.  The observed Kiessig fringes correspond to the 45 nm thickness Ga$_{0.93}$Mn$_{0.07}$As layer.  Below $T_C$, significant dichroism is seen in the reflectivity.

In figure~\ref{fig:GMA} b) we present measurements of x-ray absorption and x-ray reflectivity at a fixed angle at a temperature below $T_C$, along with the dichroism for each measurement ($LCP - RCP$).  The x-ray absorption was measured using the total fluorescence yield using the total counts on the MCP detector.  The results are consistent with previous XMCD on (Ga,Mn)As. \cite{Edmonds04}

\section{Conclusion}

We have developed a UHV compatible, 4-circle diffractometer for resonant soft x-ray scattering and reflectivity measurements.  The instrument uses in-vacuum stepper motors and motion stages to provide 9 in vacuum motions for the sample and detector positions.  The system also includes capabilities for cooling the sample to 18 K using a closed-cycle cryostat.  Detector options include a 2D sensitive channelplate detector, a photodiode detector and a single photon sensitive channeltron detector.  Initial measurements of resonant scattering in stripe ordered Nd-doped LSCO indicate reliable measurements of weak superlattice reflections are possible with this instrument.  In addition, XMCD measurements on Ni metal and magnetic resonant reflectivity measurements on (Ga,Mn)As thin films are demonstrated.

\begin{acknowledgments}
This work was supported by the BCKDF, the British Columbia Synchrotron Institute, a Canada Research Chair (G. A. S.), the Canadian Foundation for Innovation and the Natural Sciences and Engineering Research Council of Canada.  H.W. is supported by the Japan Society for the Promotion
of Science (JSPS) through its Funding Program for World-Leading Innovative R\&D on Science and Technology (FIRST Program).  J.G. gratefully acknowledges the financial support by the DFG.  We acknowledge Christian Sch{\"u}{\ss}ler-Langeheine for inspiration in the design of the sample holder. The research described in this paper was performed at the Canadian Light Source, which is supported by NSERC, NRC, CIHR, and the University of Saskatchewan.  
\end{acknowledgments}


\begin{thebibliography}{20}%
\makeatletter
\providecommand \@ifxundefined [1]{%
 \@ifx{#1\undefined}
}%
\providecommand \@ifnum [1]{%
 \ifnum #1\expandafter \@firstoftwo
 \else \expandafter \@secondoftwo
 \fi
}%
\providecommand \@ifx [1]{%
 \ifx #1\expandafter \@firstoftwo
 \else \expandafter \@secondoftwo
 \fi
}%
\providecommand \natexlab [1]{#1}%
\providecommand \enquote  [1]{``#1''}%
\providecommand \bibnamefont  [1]{#1}%
\providecommand \bibfnamefont [1]{#1}%
\providecommand \citenamefont [1]{#1}%
\providecommand \href@noop [0]{\@secondoftwo}%
\providecommand \href [0]{\begingroup \@sanitize@url \@href}%
\providecommand \@href[1]{\@@startlink{#1}\@@href}%
\providecommand \@@href[1]{\endgroup#1\@@endlink}%
\providecommand \@sanitize@url [0]{\catcode `\\12\catcode `\$12\catcode
  `\&12\catcode `\#12\catcode `\^12\catcode `\_12\catcode `\%12\relax}%
\providecommand \@@startlink[1]{}%
\providecommand \@@endlink[0]{}%
\providecommand \url  [0]{\begingroup\@sanitize@url \@url }%
\providecommand \@url [1]{\endgroup\@href {#1}{\urlprefix }}%
\providecommand \urlprefix  [0]{URL }%
\providecommand \Eprint [0]{\href }%
\@ifxundefined \urlstyle {%
  \providecommand \doi  [0]{\begingroup \@sanitize@url \@doi}%
  \providecommand \@doi [1]{\endgroup \@@startlink {\doibase
  #1}doi:\discretionary {}{}{}#1\@@endlink }%
}{%
  \providecommand \doi  [0]{doi:\discretionary{}{}{}\begingroup
  \urlstyle{rm}\Url }%
}%
\providecommand \doibase [0]{http://dx.doi.org/}%
\providecommand \Doi [0]{\begingroup \@sanitize@url \@Doi }%
\providecommand \@Doi  [1]{\endgroup\@@startlink{\doibase#1}\@@Doi}%
\providecommand \@@Doi [1]{#1\@@endlink}%
\providecommand \selectlanguage [0]{\@gobble}%
\providecommand \bibinfo  [0]{\@secondoftwo}%
\providecommand \bibfield  [0]{\@secondoftwo}%
\providecommand \translation [1]{[#1]}%
\providecommand \BibitemOpen [0]{}%
\providecommand \bibitemStop [0]{}%
\providecommand \bibitemNoStop [0]{.\EOS\space}%
\providecommand \EOS [0]{\spacefactor3000\relax}%
\providecommand \BibitemShut  [1]{\csname bibitem#1\endcsname}%
\bibitem [{\citenamefont {Abbamonte}\ \emph
  {et~al.}(2005){\natexlab{a}}\citenamefont {Abbamonte}, \citenamefont
  {Blumberg}, \citenamefont {Rusydi}, \citenamefont {Gozar}, \citenamefont
  {Evans}, \citenamefont {Siegrist}, \citenamefont {Venema}, \citenamefont
  {Eisaki}, \citenamefont {Isaacs},\ and\ \citenamefont
  {Sawatzky}}]{Abbamonte04}%
  \BibitemOpen
  \bibfield  {author} {\bibinfo {author} {\bibfnamefont {P.}~\bibnamefont
  {Abbamonte}}, \bibinfo {author} {\bibfnamefont {G.}~\bibnamefont {Blumberg}},
  \bibinfo {author} {\bibfnamefont {A.}~\bibnamefont {Rusydi}}, \bibinfo
  {author} {\bibfnamefont {A.}~\bibnamefont {Gozar}}, \bibinfo {author}
  {\bibfnamefont {P.~G.}\ \bibnamefont {Evans}}, \bibinfo {author}
  {\bibfnamefont {T.}~\bibnamefont {Siegrist}}, \bibinfo {author}
  {\bibfnamefont {L.}~\bibnamefont {Venema}}, \bibinfo {author} {\bibfnamefont
  {H.}~\bibnamefont {Eisaki}}, \bibinfo {author} {\bibfnamefont {E.~D.}\
  \bibnamefont {Isaacs}}, \ and\ \bibinfo {author} {\bibfnamefont {G.~A.}\
  \bibnamefont {Sawatzky}},\ }\href@noop {} {\bibfield  {journal} {\bibinfo
  {journal} {Nature},\ }\textbf {\bibinfo {volume} {431}},\ \bibinfo {pages}
  {1078} (\bibinfo {year} {2005}{\natexlab{a}})}\BibitemShut {NoStop}%
\bibitem [{\citenamefont {Abbamonte}\ \emph
  {et~al.}(2005){\natexlab{b}}\citenamefont {Abbamonte}, \citenamefont
  {Rusydi}, \citenamefont {Smadici}, \citenamefont {Gu}, \citenamefont
  {Sawatzky},\ and\ \citenamefont {Feng}}]{Abbamonte05}%
  \BibitemOpen
  \bibfield  {author} {\bibinfo {author} {\bibfnamefont {P.}~\bibnamefont
  {Abbamonte}}, \bibinfo {author} {\bibfnamefont {A.}~\bibnamefont {Rusydi}},
  \bibinfo {author} {\bibfnamefont {S.}~\bibnamefont {Smadici}}, \bibinfo
  {author} {\bibfnamefont {G.~D.}\ \bibnamefont {Gu}}, \bibinfo {author}
  {\bibfnamefont {G.~A.}\ \bibnamefont {Sawatzky}}, \ and\ \bibinfo {author}
  {\bibfnamefont {D.~L.}\ \bibnamefont {Feng}},\ }\href@noop {} {\bibfield
  {journal} {\bibinfo  {journal} {Nat. Phys.},\ }\textbf {\bibinfo {volume}
  {1}},\ \bibinfo {pages} {155} (\bibinfo {year}
  {2005}{\natexlab{b}})}\BibitemShut {NoStop}%
\bibitem [{\citenamefont {Wilkins}\ \emph {et~al.}(2003)\citenamefont
  {Wilkins}, \citenamefont {Spencer}, \citenamefont {Hatton}, \citenamefont
  {Collins}, \citenamefont {Roper}, \citenamefont {Prabhakaran},\ and\
  \citenamefont {Boothroyd}}]{Wilkins03}%
  \BibitemOpen
  \bibfield  {author} {\bibinfo {author} {\bibfnamefont {S.~B.}\ \bibnamefont
  {Wilkins}}, \bibinfo {author} {\bibfnamefont {P.~D.}\ \bibnamefont
  {Spencer}}, \bibinfo {author} {\bibfnamefont {P.~D.}\ \bibnamefont {Hatton}},
  \bibinfo {author} {\bibfnamefont {S.~P.}\ \bibnamefont {Collins}}, \bibinfo
  {author} {\bibfnamefont {M.~D.}\ \bibnamefont {Roper}}, \bibinfo {author}
  {\bibfnamefont {D.}~\bibnamefont {Prabhakaran}}, \ and\ \bibinfo {author}
  {\bibfnamefont {A.~T.}\ \bibnamefont {Boothroyd}},\ }\href@noop {} {\bibfield
   {journal} {\bibinfo  {journal} {Phys. Rev. Lett.},\ }\textbf {\bibinfo
  {volume} {91}},\ \bibinfo {pages} {167205} (\bibinfo {year}
  {2003})}\BibitemShut {NoStop}%
\bibitem [{\citenamefont {Thomas}\ \emph {et~al.}(2004)\citenamefont {Thomas},
  \citenamefont {Hill}, \citenamefont {Grenier}, \citenamefont {Kim},
  \citenamefont {Abbamonte}, \citenamefont {Venema}, \citenamefont {Rusydi},
  \citenamefont {Tomioka}, \citenamefont {Tokura}, \citenamefont {McMorrow},
  \citenamefont {Sawatzky},\ and\ \citenamefont {van Veenendaal}}]{Thomas04}%
  \BibitemOpen
  \bibfield  {author} {\bibinfo {author} {\bibfnamefont {K.~J.}\ \bibnamefont
  {Thomas}}, \bibinfo {author} {\bibfnamefont {J.~P.}\ \bibnamefont {Hill}},
  \bibinfo {author} {\bibfnamefont {S.}~\bibnamefont {Grenier}}, \bibinfo
  {author} {\bibfnamefont {Y.-J.}\ \bibnamefont {Kim}}, \bibinfo {author}
  {\bibfnamefont {P.}~\bibnamefont {Abbamonte}}, \bibinfo {author}
  {\bibfnamefont {L.}~\bibnamefont {Venema}}, \bibinfo {author} {\bibfnamefont
  {A.}~\bibnamefont {Rusydi}}, \bibinfo {author} {\bibfnamefont
  {Y.}~\bibnamefont {Tomioka}}, \bibinfo {author} {\bibfnamefont
  {Y.}~\bibnamefont {Tokura}}, \bibinfo {author} {\bibfnamefont {D.~F.}\
  \bibnamefont {McMorrow}}, \bibinfo {author} {\bibfnamefont {G.}~\bibnamefont
  {Sawatzky}}, \ and\ \bibinfo {author} {\bibfnamefont {M.}~\bibnamefont {van
  Veenendaal}},\ }\href@noop {} {\bibfield  {journal} {\bibinfo  {journal}
  {Phys. Rev. Lett.},\ }\textbf {\bibinfo {volume} {92}},\ \bibinfo {pages}
  {231204} (\bibinfo {year} {2004})}\BibitemShut {NoStop}%
\bibitem [{\citenamefont {Sch{\"u}{\ss}ler-Langeheine}\ \emph
  {et~al.}(2005)\citenamefont {Sch{\"u}{\ss}ler-Langeheine}, \citenamefont
  {Schlappa}, \citenamefont {Tanaka}, \citenamefont {Hu}, \citenamefont
  {Chang}, \citenamefont {Schierle}, \citenamefont {Benomar}, \citenamefont
  {Ott}, \citenamefont {Weschke}, \citenamefont {Kaindl}, \citenamefont
  {Friedt}, \citenamefont {Sawatzky}, \citenamefont {Lin}, \citenamefont
  {Chen}, \citenamefont {Braden},\ and\ \citenamefont {Tjeng}}]{Schusler05}%
  \BibitemOpen
  \bibfield  {author} {\bibinfo {author} {\bibfnamefont {C.}~\bibnamefont
  {Sch{\"u}{\ss}ler-Langeheine}}, \bibinfo {author} {\bibfnamefont
  {J.}~\bibnamefont {Schlappa}}, \bibinfo {author} {\bibfnamefont
  {A.}~\bibnamefont {Tanaka}}, \bibinfo {author} {\bibfnamefont
  {Z.}~\bibnamefont {Hu}}, \bibinfo {author} {\bibfnamefont {C.~F.}\
  \bibnamefont {Chang}}, \bibinfo {author} {\bibfnamefont {E.}~\bibnamefont
  {Schierle}}, \bibinfo {author} {\bibfnamefont {M.}~\bibnamefont {Benomar}},
  \bibinfo {author} {\bibfnamefont {H.}~\bibnamefont {Ott}}, \bibinfo {author}
  {\bibfnamefont {E.}~\bibnamefont {Weschke}}, \bibinfo {author} {\bibfnamefont
  {G.}~\bibnamefont {Kaindl}}, \bibinfo {author} {\bibfnamefont
  {O.}~\bibnamefont {Friedt}}, \bibinfo {author} {\bibfnamefont {G.~A.}\
  \bibnamefont {Sawatzky}}, \bibinfo {author} {\bibfnamefont {H.-J.}\
  \bibnamefont {Lin}}, \bibinfo {author} {\bibfnamefont {C.~T.}\ \bibnamefont
  {Chen}}, \bibinfo {author} {\bibfnamefont {M.}~\bibnamefont {Braden}}, \ and\
  \bibinfo {author} {\bibfnamefont {L.~H.}\ \bibnamefont {Tjeng}},\ }\href@noop
  {} {\bibfield  {journal} {\bibinfo  {journal} {Phys. Rev. Lett.},\ }\textbf
  {\bibinfo {volume} {95}},\ \bibinfo {pages} {156402} (\bibinfo {year}
  {2005})}\BibitemShut {NoStop}%
\bibitem [{\citenamefont {Huang}\ \emph {et~al.}(2006)\citenamefont {Huang},
  \citenamefont {Lin}, \citenamefont {Okamoto}, \citenamefont {Chao},
  \citenamefont {Jeng}, \citenamefont {Guo}, \citenamefont {Hsu}, \citenamefont
  {Huang}, \citenamefont {Ling}, \citenamefont {Wu}, \citenamefont {Yang},\
  and\ \citenamefont {Chen}}]{Huang06}%
  \BibitemOpen
  \bibfield  {author} {\bibinfo {author} {\bibfnamefont {D.~J.}\ \bibnamefont
  {Huang}}, \bibinfo {author} {\bibfnamefont {H.-J.}\ \bibnamefont {Lin}},
  \bibinfo {author} {\bibfnamefont {J.}~\bibnamefont {Okamoto}}, \bibinfo
  {author} {\bibfnamefont {K.~S.}\ \bibnamefont {Chao}}, \bibinfo {author}
  {\bibfnamefont {H.-T.}\ \bibnamefont {Jeng}}, \bibinfo {author}
  {\bibfnamefont {G.~Y.}\ \bibnamefont {Guo}}, \bibinfo {author} {\bibfnamefont
  {C.-H.}\ \bibnamefont {Hsu}}, \bibinfo {author} {\bibfnamefont {C.-M.}\
  \bibnamefont {Huang}}, \bibinfo {author} {\bibfnamefont {D.~C.}\ \bibnamefont
  {Ling}}, \bibinfo {author} {\bibfnamefont {W.~B.}\ \bibnamefont {Wu}},
  \bibinfo {author} {\bibfnamefont {C.~S.}\ \bibnamefont {Yang}}, \ and\
  \bibinfo {author} {\bibfnamefont {C.~T.}\ \bibnamefont {Chen}},\ }\href@noop
  {} {\bibfield  {journal} {\bibinfo  {journal} {Phys. Rev. Lett.},\ }\textbf
  {\bibinfo {volume} {96}},\ \bibinfo {pages} {096401} (\bibinfo {year}
  {2006})}\BibitemShut {NoStop}%
\bibitem [{\citenamefont {Beale}\ \emph {et~al.}(2010)\citenamefont {Beale},
  \citenamefont {Hase}, \citenamefont {Iida}, \citenamefont {Endo},
  \citenamefont {Steadman}, \citenamefont {Marshall}, \citenamefont {Dhesi},
  \citenamefont {van~der Laan},\ and\ \citenamefont {Hatton}}]{Beale10}%
  \BibitemOpen
  \bibfield  {author} {\bibinfo {author} {\bibfnamefont {T.~A.~W.}\
  \bibnamefont {Beale}}, \bibinfo {author} {\bibfnamefont {T.~P.~A.}\
  \bibnamefont {Hase}}, \bibinfo {author} {\bibfnamefont {T.}~\bibnamefont
  {Iida}}, \bibinfo {author} {\bibfnamefont {K.}~\bibnamefont {Endo}}, \bibinfo
  {author} {\bibfnamefont {P.}~\bibnamefont {Steadman}}, \bibinfo {author}
  {\bibfnamefont {A.~R.}\ \bibnamefont {Marshall}}, \bibinfo {author}
  {\bibfnamefont {S.~S.}\ \bibnamefont {Dhesi}}, \bibinfo {author}
  {\bibfnamefont {G.}~\bibnamefont {van~der Laan}}, \ and\ \bibinfo {author}
  {\bibfnamefont {P.~D.}\ \bibnamefont {Hatton}},\ }\href@noop {} {\bibfield
  {journal} {\bibinfo  {journal} {Rev. Sci. Instrum.},\ }\textbf {\bibinfo
  {volume} {81}},\ \bibinfo {pages} {073904} (\bibinfo {year}
  {2010})}\BibitemShut {NoStop}%
\bibitem [{\citenamefont {Br{\"u}ck}\ \emph {et~al.}(2008)\citenamefont
  {Br{\"u}ck}, \citenamefont {Bauknecht}, \citenamefont {Ludescher},
  \citenamefont {Goering},\ and\ \citenamefont {Sch{\"u}tz}}]{Bruck08}%
  \BibitemOpen
  \bibfield  {author} {\bibinfo {author} {\bibfnamefont {S.}~\bibnamefont
  {Br{\"u}ck}}, \bibinfo {author} {\bibfnamefont {S.}~\bibnamefont
  {Bauknecht}}, \bibinfo {author} {\bibfnamefont {B.}~\bibnamefont
  {Ludescher}}, \bibinfo {author} {\bibfnamefont {E.}~\bibnamefont {Goering}},
  \ and\ \bibinfo {author} {\bibfnamefont {G.}~\bibnamefont {Sch{\"u}tz}},\
  }\href@noop {} {\bibfield  {journal} {\bibinfo  {journal} {Rev. Sci.
  Instrum.},\ }\textbf {\bibinfo {volume} {79}},\ \bibinfo {pages} {083109}
  (\bibinfo {year} {2008})}\BibitemShut {NoStop}%
\bibitem [{\citenamefont {Grabis}\ \emph {et~al.}(2003)\citenamefont {Grabis},
  \citenamefont {Nefedov},\ and\ \citenamefont {Zabel}}]{Grabis03}%
  \BibitemOpen
  \bibfield  {author} {\bibinfo {author} {\bibfnamefont {J.}~\bibnamefont
  {Grabis}}, \bibinfo {author} {\bibfnamefont {A.}~\bibnamefont {Nefedov}}, \
  and\ \bibinfo {author} {\bibfnamefont {H.}~\bibnamefont {Zabel}},\
  }\href@noop {} {\bibfield  {journal} {\bibinfo  {journal} {Rev. Sci.
  Instrum.},\ }\textbf {\bibinfo {volume} {74}},\ \bibinfo {pages} {4048}
  (\bibinfo {year} {2003})}\BibitemShut {NoStop}%
\bibitem [{\citenamefont {Roper}\ \emph {et~al.}(2001)\citenamefont {Roper},
  \citenamefont {van~der Laan}, \citenamefont {D{\"u}rr}, \citenamefont
  {Dudzik}, \citenamefont {Collins}, \citenamefont {Miller},\ and\
  \citenamefont {Thompson}}]{Roper01}%
  \BibitemOpen
  \bibfield  {author} {\bibinfo {author} {\bibfnamefont {M.}~\bibnamefont
  {Roper}}, \bibinfo {author} {\bibfnamefont {G.}~\bibnamefont {van~der Laan}},
  \bibinfo {author} {\bibfnamefont {H.}~\bibnamefont {D{\"u}rr}}, \bibinfo
  {author} {\bibfnamefont {E.}~\bibnamefont {Dudzik}}, \bibinfo {author}
  {\bibfnamefont {S.}~\bibnamefont {Collins}}, \bibinfo {author} {\bibfnamefont
  {M.}~\bibnamefont {Miller}}, \ and\ \bibinfo {author} {\bibfnamefont
  {S.}~\bibnamefont {Thompson}},\ }\href@noop {} {\bibfield  {journal}
  {\bibinfo  {journal} {Nucl. Instrum. Methods Phys. Res. A},\ }\textbf
  {\bibinfo {volume} {467 - 468}},\ \bibinfo {pages} {1101} (\bibinfo {year}
  {2001})}\BibitemShut {NoStop}%
\bibitem [{\citenamefont {Yu}\ \emph {et~al.}(2005)\citenamefont {Yu},
  \citenamefont {Wilhelmi}, \citenamefont {Moser}, \citenamefont {Vidyaraj},
  \citenamefont {Gao}, \citenamefont {Wee}, \citenamefont {Nyunt},
  \citenamefont {Qian},\ and\ \citenamefont {Zheng}}]{Yu05}%
  \BibitemOpen
  \bibfield  {author} {\bibinfo {author} {\bibfnamefont {X.}~\bibnamefont
  {Yu}}, \bibinfo {author} {\bibfnamefont {O.}~\bibnamefont {Wilhelmi}},
  \bibinfo {author} {\bibfnamefont {H.~O.}\ \bibnamefont {Moser}}, \bibinfo
  {author} {\bibfnamefont {S.~V.}\ \bibnamefont {Vidyaraj}}, \bibinfo {author}
  {\bibfnamefont {X.}~\bibnamefont {Gao}}, \bibinfo {author} {\bibfnamefont
  {A.~T.}\ \bibnamefont {Wee}}, \bibinfo {author} {\bibfnamefont
  {T.}~\bibnamefont {Nyunt}}, \bibinfo {author} {\bibfnamefont
  {H.}~\bibnamefont {Qian}}, \ and\ \bibinfo {author} {\bibfnamefont
  {H.}~\bibnamefont {Zheng}},\ }\href@noop {} {\bibfield  {journal} {\bibinfo
  {journal} {J. Electron Spectrosc. Relat. Phenom.},\ }\textbf {\bibinfo
  {volume} {144-147}},\ \bibinfo {pages} {1031} (\bibinfo {year}
  {2005})}\BibitemShut {NoStop}%
\bibitem [{\citenamefont {Takeuchi}\ \emph {et~al.}(2009)\citenamefont
  {Takeuchi}, \citenamefont {Chainani}, \citenamefont {Takata}, \citenamefont
  {Tanaka}, \citenamefont {Oura}, \citenamefont {Tsubota}, \citenamefont
  {Senba}, \citenamefont {Ohashi}, \citenamefont {Mochiku}, \citenamefont
  {Hirata},\ and\ \citenamefont {Shin}}]{Takeuchi09}%
  \BibitemOpen
  \bibfield  {author} {\bibinfo {author} {\bibfnamefont {T.}~\bibnamefont
  {Takeuchi}}, \bibinfo {author} {\bibfnamefont {A.}~\bibnamefont {Chainani}},
  \bibinfo {author} {\bibfnamefont {Y.}~\bibnamefont {Takata}}, \bibinfo
  {author} {\bibfnamefont {Y.}~\bibnamefont {Tanaka}}, \bibinfo {author}
  {\bibfnamefont {M.}~\bibnamefont {Oura}}, \bibinfo {author} {\bibfnamefont
  {M.}~\bibnamefont {Tsubota}}, \bibinfo {author} {\bibfnamefont
  {Y.}~\bibnamefont {Senba}}, \bibinfo {author} {\bibfnamefont
  {H.}~\bibnamefont {Ohashi}}, \bibinfo {author} {\bibfnamefont
  {T.}~\bibnamefont {Mochiku}}, \bibinfo {author} {\bibfnamefont
  {K.}~\bibnamefont {Hirata}}, \ and\ \bibinfo {author} {\bibfnamefont
  {S.}~\bibnamefont {Shin}},\ }\href@noop {} {\bibfield  {journal} {\bibinfo
  {journal} {Rev. Sci. Instrum.},\ }\textbf {\bibinfo {volume} {80}},\ \bibinfo
  {pages} {023905} (\bibinfo {year} {2009})}\BibitemShut {NoStop}%
\bibitem [{\citenamefont {Henke}\ \emph {et~al.}(1993)\citenamefont {Henke},
  \citenamefont {Gullikson},\ and\ \citenamefont {Davis}}]{Henke93}%
  \BibitemOpen
  \bibfield  {author} {\bibinfo {author} {\bibfnamefont {B.}~\bibnamefont
  {Henke}}, \bibinfo {author} {\bibfnamefont {E.}~\bibnamefont {Gullikson}}, \
  and\ \bibinfo {author} {\bibfnamefont {J.}~\bibnamefont {Davis}},\
  }\href@noop {} {\bibfield  {journal} {\bibinfo  {journal} {At. Data Nucl.
  Data Tables},\ }\textbf {\bibinfo {volume} {54}},\ \bibinfo {pages} {181}
  (\bibinfo {year} {1993})}\BibitemShut {NoStop}%
\bibitem [{\citenamefont {Chen}\ \emph {et~al.}(1990)\citenamefont {Chen},
  \citenamefont {Sette}, \citenamefont {Ma},\ and\ \citenamefont
  {Modesti}}]{Chen90}%
  \BibitemOpen
  \bibfield  {author} {\bibinfo {author} {\bibfnamefont {C.~T.}\ \bibnamefont
  {Chen}}, \bibinfo {author} {\bibfnamefont {F.}~\bibnamefont {Sette}},
  \bibinfo {author} {\bibfnamefont {Y.}~\bibnamefont {Ma}}, \ and\ \bibinfo
  {author} {\bibfnamefont {S.}~\bibnamefont {Modesti}},\ }\Doi
  {10.1103/PhysRevB.42.7262} {\bibfield  {journal} {\bibinfo  {journal} {Phys.
  Rev. B},\ }\textbf {\bibinfo {volume} {42}},\ \bibinfo {pages} {7262}
  (\bibinfo {year} {1990})}\BibitemShut {NoStop}%
\bibitem [{\citenamefont {Tranquada}\ \emph {et~al.}(1995)\citenamefont
  {Tranquada}, \citenamefont {Sternlieb}, \citenamefont {Axe}, \citenamefont
  {Nakamura},\ and\ \citenamefont {Uchida}}]{Tranquada95}%
  \BibitemOpen
  \bibfield  {author} {\bibinfo {author} {\bibfnamefont {J.~M.}\ \bibnamefont
  {Tranquada}}, \bibinfo {author} {\bibfnamefont {B.~J.}\ \bibnamefont
  {Sternlieb}}, \bibinfo {author} {\bibfnamefont {J.~D.}\ \bibnamefont {Axe}},
  \bibinfo {author} {\bibfnamefont {Y.}~\bibnamefont {Nakamura}}, \ and\
  \bibinfo {author} {\bibfnamefont {S.}~\bibnamefont {Uchida}},\ }\href@noop {}
  {\bibfield  {journal} {\bibinfo  {journal} {Nature},\ }\textbf {\bibinfo
  {volume} {375}},\ \bibinfo {pages} {561} (\bibinfo {year}
  {1995})}\BibitemShut {NoStop}%
\bibitem [{\citenamefont {Fink}\ \emph {et~al.}(2009)\citenamefont {Fink},
  \citenamefont {Schierle}, \citenamefont {Weschke}, \citenamefont {Geck},
  \citenamefont {Hawthorn}, \citenamefont {Soltwisch}, \citenamefont {Wadati},
  \citenamefont {Wu}, \citenamefont {Durr}, \citenamefont {Wizent},
  \citenamefont {Buchner},\ and\ \citenamefont {Sawatzky}}]{Fink09}%
  \BibitemOpen
  \bibfield  {author} {\bibinfo {author} {\bibfnamefont {J.}~\bibnamefont
  {Fink}}, \bibinfo {author} {\bibfnamefont {E.}~\bibnamefont {Schierle}},
  \bibinfo {author} {\bibfnamefont {E.}~\bibnamefont {Weschke}}, \bibinfo
  {author} {\bibfnamefont {J.}~\bibnamefont {Geck}}, \bibinfo {author}
  {\bibfnamefont {D.}~\bibnamefont {Hawthorn}}, \bibinfo {author}
  {\bibfnamefont {V.}~\bibnamefont {Soltwisch}}, \bibinfo {author}
  {\bibfnamefont {H.}~\bibnamefont {Wadati}}, \bibinfo {author} {\bibfnamefont
  {H.-H.}\ \bibnamefont {Wu}}, \bibinfo {author} {\bibfnamefont {H.~A.}\
  \bibnamefont {Durr}}, \bibinfo {author} {\bibfnamefont {N.}~\bibnamefont
  {Wizent}}, \bibinfo {author} {\bibfnamefont {B.}~\bibnamefont {Buchner}}, \
  and\ \bibinfo {author} {\bibfnamefont {G.~A.}\ \bibnamefont {Sawatzky}},\
  }\Doi {10.1103/PhysRevB.79.100502} {\bibfield  {journal} {\bibinfo  {journal}
  {Phys. Rev. B},\ }\textbf {\bibinfo {volume} {79}},\ \bibinfo {eid} {100502}
  (\bibinfo {year} {2009})}\BibitemShut {NoStop}%
\bibitem [{\citenamefont {Chen}\ \emph {et~al.}(1992)\citenamefont {Chen},
  \citenamefont {Tjeng}, \citenamefont {Kwo}, \citenamefont {Kao},
  \citenamefont {Rudolf}, \citenamefont {Sette},\ and\ \citenamefont
  {Fleming}}]{Chen92}%
  \BibitemOpen
  \bibfield  {author} {\bibinfo {author} {\bibfnamefont {C.~T.}\ \bibnamefont
  {Chen}}, \bibinfo {author} {\bibfnamefont {L.~H.}\ \bibnamefont {Tjeng}},
  \bibinfo {author} {\bibfnamefont {J.}~\bibnamefont {Kwo}}, \bibinfo {author}
  {\bibfnamefont {H.~L.}\ \bibnamefont {Kao}}, \bibinfo {author} {\bibfnamefont
  {P.}~\bibnamefont {Rudolf}}, \bibinfo {author} {\bibfnamefont
  {F.}~\bibnamefont {Sette}}, \ and\ \bibinfo {author} {\bibfnamefont {R.~M.}\
  \bibnamefont {Fleming}},\ }\Doi {10.1103/PhysRevLett.68.2543} {\bibfield
  {journal} {\bibinfo  {journal} {Phys. Rev. Lett.},\ }\textbf {\bibinfo
  {volume} {68}},\ \bibinfo {pages} {2543} (\bibinfo {year}
  {1992})}\BibitemShut {NoStop}%
\bibitem [{\citenamefont {Tranquada}\ \emph {et~al.}(1996)\citenamefont
  {Tranquada}, \citenamefont {Axe}, \citenamefont {Ichikawa}, \citenamefont
  {Nakamura}, \citenamefont {Uchida},\ and\ \citenamefont
  {Nachumi}}]{Tranquada96}%
  \BibitemOpen
  \bibfield  {author} {\bibinfo {author} {\bibfnamefont {J.~M.}\ \bibnamefont
  {Tranquada}}, \bibinfo {author} {\bibfnamefont {J.~D.}\ \bibnamefont {Axe}},
  \bibinfo {author} {\bibfnamefont {N.}~\bibnamefont {Ichikawa}}, \bibinfo
  {author} {\bibfnamefont {Y.}~\bibnamefont {Nakamura}}, \bibinfo {author}
  {\bibfnamefont {S.}~\bibnamefont {Uchida}}, \ and\ \bibinfo {author}
  {\bibfnamefont {B.}~\bibnamefont {Nachumi}},\ }\href@noop {} {\bibfield
  {journal} {\bibinfo  {journal} {Phys. Rev. B},\ }\textbf {\bibinfo {volume}
  {54}},\ \bibinfo {pages} {7489} (\bibinfo {year} {1996})}\BibitemShut
  {NoStop}%
\bibitem [{\citenamefont {Eisebitt}\ \emph {et~al.}(1993)\citenamefont
  {Eisebitt}, \citenamefont {B\"oske}, \citenamefont {Rubensson},\ and\
  \citenamefont {Eberhardt}}]{Eisebitt93}%
  \BibitemOpen
  \bibfield  {author} {\bibinfo {author} {\bibfnamefont {S.}~\bibnamefont
  {Eisebitt}}, \bibinfo {author} {\bibfnamefont {T.}~\bibnamefont {B\"oske}},
  \bibinfo {author} {\bibfnamefont {J.-E.}\ \bibnamefont {Rubensson}}, \ and\
  \bibinfo {author} {\bibfnamefont {W.}~\bibnamefont {Eberhardt}},\ }\Doi
  {10.1103/PhysRevB.47.14103} {\bibfield  {journal} {\bibinfo  {journal} {Phys.
  Rev. B},\ }\textbf {\bibinfo {volume} {47}},\ \bibinfo {pages} {14103}
  (\bibinfo {year} {1993})}\BibitemShut {NoStop}%
\bibitem [{\citenamefont {Edmonds}\ \emph {et~al.}(2004)\citenamefont
  {Edmonds}, \citenamefont {Farley}, \citenamefont {Campion}, \citenamefont
  {Foxon}, \citenamefont {Gallagher}, \citenamefont {Johal}, \citenamefont
  {van~der Laan}, \citenamefont {MacKenzie}, \citenamefont {Chapman},\ and\
  \citenamefont {Arenholz}}]{Edmonds04}%
  \BibitemOpen
  \bibfield  {author} {\bibinfo {author} {\bibfnamefont {K.~W.}\ \bibnamefont
  {Edmonds}}, \bibinfo {author} {\bibfnamefont {N.~R.~S.}\ \bibnamefont
  {Farley}}, \bibinfo {author} {\bibfnamefont {R.~P.}\ \bibnamefont {Campion}},
  \bibinfo {author} {\bibfnamefont {C.~T.}\ \bibnamefont {Foxon}}, \bibinfo
  {author} {\bibfnamefont {B.~L.}\ \bibnamefont {Gallagher}}, \bibinfo {author}
  {\bibfnamefont {T.~K.}\ \bibnamefont {Johal}}, \bibinfo {author}
  {\bibfnamefont {G.}~\bibnamefont {van~der Laan}}, \bibinfo {author}
  {\bibfnamefont {M.}~\bibnamefont {MacKenzie}}, \bibinfo {author}
  {\bibfnamefont {J.~N.}\ \bibnamefont {Chapman}}, \ and\ \bibinfo {author}
  {\bibfnamefont {E.}~\bibnamefont {Arenholz}},\ }\href@noop {} {\bibfield
  {journal} {\bibinfo  {journal} {Appl. Phys. Lett.},\ }\textbf {\bibinfo
  {volume} {84}},\ \bibinfo {pages} {4065} (\bibinfo {year}
  {2004})}\BibitemShut {NoStop}%
\end{thebibliography}
\end{document}